\documentclass{emulateapj}
\usepackage{epsfig}
\usepackage{rotating}
\usepackage[utf8]{inputenc}
\usepackage{lineno}
\usepackage{natbib} 
\usepackage{graphicx}
\usepackage{xcolor} 
\usepackage{babel} 
\newcommand{\rpm}{\raisebox{.2ex}{$\scriptstyle\pm$}} 
\usepackage{amsmath}
\usepackage[flushleft]{threeparttable}

\def\HII{\hbox{H\,{\sc ii}}}

\def\arcsec{\hbox{$^{\prime\prime}$}}

\shorttitle{The Radio Emission from Radiative Filaments of Cygnus Loop}
\shortauthors{D. Uro{\v s}evi{\' c} et al.}

\begin{document}

\title{The Radio Emission from Radiative Filaments of Cygnus Loop}
\author{D. Uro{\v s}evi{\' c$^{1}$}, M. Andjeli{\' c}$^{1}$, M. D. Filipovi{\' c}$^{2}$, Z. J. Smeaton$^{2}$, E. Crawford$^{2}$, J. Raymond$^{3}$, D. Oni{\' c}$^{1}$}
\affil{$^{1}$Department of Astronomy, Faculty of Mathematics,
University of Belgrade, Studentski trg 16, 11000 Belgrade, Serbia}
\affil{$^{2}$Western Sydney University, Locked Bag 1797, Penrith South DC, NSW 2751, Australia}
\affil{$^{3}$Harvard-Smithsonian Center for Astrophysics, 60 Garden St., Cambridge, MA 02138, USA}
\email{dejanu@math.rs}

\begin{abstract}
The Galactic supernova remnant (SNR) Cygnus Loop emerges as an ideal laboratory for analyzing the different radiation mechanisms, as well as the particle acceleration mechanisms at different types of shocks. In order to determine radio spectral indices of non-radiative and radiative filaments in Cygnus Loop, we observed previously optically analyzed filaments with the  Karl G. Jansky Very Large Array (VLA). At 1 and 5\,GHz, we detected only radiative filaments in the field of view. Non-radiative optical filaments are also present, but were not detected in radio. Contrary to the expected non-thermal spectral slopes characteristic of SNRs, we instead observed spectral slopes characteristic of the thermal radiation mechanism from the radiative filaments in Cygnus Loop. These evolutionary older parts of Cygnus Loop radiate at radio frequencies predominantly via the thermal bremsstrahlung mechanism, and in that sense their emission more closely resembles the radio emission of \HII\ regions rather than the radio emission of SNRs.

\end{abstract}


\section{Introduction}
\noindent   \color{black}

 {Shell-type supernova remnants (SNRs) are recognized as diffuse sources of radio continuum emission, predominantly through the synchrotron mechanism, both within the Milky Way and in several nearby galaxies \citep{2012SSRv..166..231R, 2014SerAJ.189...15G, 2017ApJS..230....2B,2019A&A...631A.127M,2023MNRAS.518.2574B,2023ApJS..265...53R,2024MNRAS.529.2443C}. Charged particles are accelerated at the shock waves via Fermi processes, making SNRs important sources of Galactic cosmic-rays. There are other mechanisms of radiation contributing to SNR continuum emission in the X and $\gamma$-ray domains, as well as emission lines in the IR, optical and UV-domains, that can be used to determine electron temperature, density and other properties of disturbed matter by the SNR shock waves \citep[see][for more details]{vink12,2021map..book.....F}. In addition, the radio continuum from evolved SNRs, especially those expanding in the high density environments, can include a significant thermal bremsstrahlung component alongside the non-thermal synchrotron emission (Uro{\v s}evi{\' c} \& Pannuti 2005; Uro{\v s}evi{\' c} et al.~2007; Oni{\' c} et al.~2012; Oni{\'c} 2013; Uro{\v s}evi{\' c} 2014).}

 {The dominant charged particle acceleration mechanism at SNR shocks is first-order Fermi acceleration, in particular, the so-called diffusive shock acceleration (DSA) process. In the test-particle limit for strong shocks with compression ratio $r=4$, DSA predicts a particle energy spectrum $N(E)\propto E^{-2}$ and, consequently, a radio continuum synchrotron spectral index $\alpha\approx -0.5$, where the flux density $S_\nu\propto\nu^\alpha$ \citep{axetal77,bo78,kry77,bell78a,1978MNRAS.182..443B,1987PhR...154....1B}. Reynolds~(2011) discussed how non-linear DSA can produce radio spectral indices that deviate from the canonical test-particle value and that can be responsible for spectral flattening at higher radio continuum frequencies. However, the revised non-linear DSA model that includes the so-called postcursor region imply spectra that are steeper than the standard DSA prediction \citep{diesing21}. In addition, second-order Fermi (Fermi~II) processes can also operate in the turbulent downstream regions, and may be responsible for some observed values of integrated radio spectral indices of $\alpha\approx (-0.5,-0.3)$ (Schlickeiser \& F{\"u}rst 1989; Ostrowski 1999).}

 {SNR shocks can be classified into two types based on their evolutionary stage and  {shock} velocity. During classical free-expansion and Sedov-Taylor phases of evolution, so-called non-radiative shocks with high velocities (typically $>250$\,km\,s$^{-1}$) efficiently accelerate particles to ultrarelativistic energies via DSA process \citep[see][and references therein]{uraron19}. These electrons produce radio synchrotron emission, as well as X-ray synchrotron and $\gamma$-ray emission through inverse Compton or non-thermal bremsstrahlung processes (Tutone et al.~2021). In addition, cosmic-ray ions may be responsible for $\gamma$-ray emission through the so-called hadronic scenario via neutral pion decay (Uchiyama et al.~2010). Non-radiative shocks can produce faint H$\alpha$ emission by collisional excitation of neutral H atoms that pass through the shock. Because of the high temperatures, this can occur at low densities, and little radio bremsstrahlung is produced. On the other hand, for evolutionary older SNRs in post Sedov-Taylor phases, with low shock velocities ($< 250\,\mathrm{km\,s^{-1}}$), the shocks become radiative. That is, the shock heats the gas, but radiative cooling by the excitation of optical and UV spectral lines cools the gas to temperatures
below $10^4$\,K. As the gas cools, it is compressed in order to maintain
pressure equilibrium. DSA becomes more and more inefficient in these slower
shocks. However, synchrotron radio emission may arise from adiabatic compression of pre-existing cosmic-ray electrons and magnetic fields (van der Laan mechanism) or from particle re-acceleration (van der Laan 1962; Uchiyama et al.~2010; Raymond et al.~2020; Uro{\v s}evi{\' c} 2022; Filipovi\'c et al.~2023, 2024). Bright H$\alpha$ emission arises from recombination in the gas that has cooled and become dense. It is accompanied
by thermal bremsstrahlung (free-free) emission depends strongly on the square of the post-shock electron number density $n_{\mathrm{e}}$ and only weakly on temperature $T_{\mathrm{e}}$, with the volume emissivity scaling mainly as $\propto n_\mathrm{e}^2 T_{\mathrm{e}}^{-0.5}$, neglecting the slowly varying Gaunt factor for this purpose. In that sense, it is plausible that evolutionary older SNRs associated with high density surroundings can exhibit both thermal bremsstrahlung and non-thermal synchrotron radio continuum emission \citep{2012ApJ...756...61O}. The corresponding radio spectral indices for thermal bremsstrahlung are between $-0.1$ and $2$ ($\alpha=-0.1$ for the totally optically thin medium; $\alpha=2$ for the totally optically thick medium).}

 {Classification of SNRs as young, middle-aged, mature, and old is conventionally based on their estimated age and  {shock} emission properties and evolutionary phase (Truelove \& McKee 1999; Vink 2012; Chousein-Basia et al.~2026). However, the actual classical dynamical evolutionary stage (free-expansion, Sedov-Taylor, and post Sedov-Taylor) depends critically on both the real age, as well as on density and structure of the surrounding ISM. For example, SNRs expanding through a low-density homogeneous ISM transition into subsequent evolutionary phases at much later times compared to those propagating through denser media. Of particular interest are remnants that expand through inhomogeneous environments or interact with molecular clouds, resulting in spatially varying local dynamical stages within the same object. Consequently, for many SNRs, including the Cygnus Loop, due to expansion through a complex, inhomogeneous environment, different parts of the same remnant can simultaneously be in different dynamical phases.}

 {The Cygnus Loop (G74.0--8.5) is a well-known Galactic SNR, whose distance is estimated to be $725\pm15$~pc (Fesen et al.~2021) based on stellar parallaxes from the Gaia data, though Ritchey, Federman \& Lambert (2024) showed that it may be as distant as $800$\,pc. It is a shell-type remnant without associated pulsar wind nebula.} This middle-aged remnant ($\sim\!\!20\,000$ years old) has an angular diameter of $\sim\!3^{\circ} $ on the sky. Its size and brightness make it suitable for studying smaller structures e.g.~filaments. Raymond et al.~(2020a,b) observed a portion of the western Cygnus Loop with the Hubble Space Telescope (HST) and a ground-based spectrograph. They were able to determine the pre-shock magnetic field and the compression ratio. By assuming that the pre-shock energetic electron and proton densities were the same as those measured outside our solar system by the Voyager satellites \citep{cummings16}, they predicted radio synchrotron and $\gamma$-ray hadronic pion emissivities based on the van der Laan mechanism in agreement with observations, though the filling factor of the emitting region remained a substantial uncertainty. Subsequently, Tutone et al.~(2021) used better FERMI $\gamma$-ray maps and the reacceleration model of Uchiyama (2010) to separate the contributions of radiative and non-radiative regions within the Cygnus Loop.  {However, interpretation of the radio emission contains another uncertainty. The optically bright radiative shocks can also produce thermal bremsstrahlung radio emission, and the variations in spectral index around the Cygnus Loop \citep{green90} could result from varying ratios of thermal to non-thermal emission.} The range in spectral index $\alpha$ among Green’s $12$ large regions was $-0.56$ to $-0.23$, which would suggest that some regions are dominated by thermal emission and others by non-thermal emission, if the spectra are optically thin. Much of the uncertainty in the interpretation of the radio emission stems from the low spatial resolution of the radio observations.

\begin{figure*}
  \centering
\includegraphics[width=12.5cm]{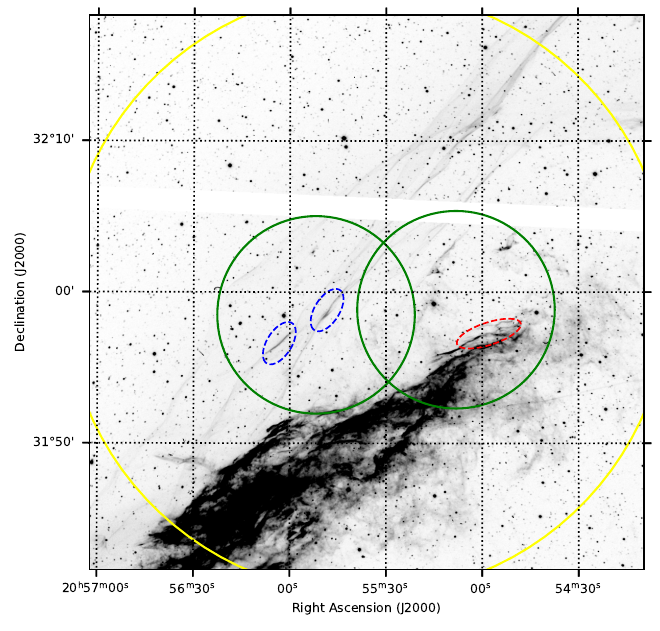}
\caption{{Context of our observations.} The background image shows H$\alpha$ emission (from 1993) of the northeastern part of Cygnus Loop (courtesy of R.~Fesen). Yellow and green circles mark  VLA pointing (primary beams) at 1~GHz and 5~GHz. Two fields of view at 5~GHz were needed in order to cover both  { non-radiative filaments (blue dashed-line ellipses), labeled with F and K in Vu{\v{c}}eti{\'c} et al.~(2023)  and radiative filaments (red dashed-line ellipse), labeled with H in Vu{\v{c}}eti{\'c} et al.~(2023).}}
\label{fig:1}
 \end{figure*}

 {The Cygnus Loop SNR contains both non-radiative as well as radiative filaments. Different authors have measured the speeds of non-radiative filaments using different techniques (e.g.~Salvesen et al.~2009, Medina et al.~2014). Recently, the speed of the shock wave have been determined by measuring the proper motions of both non-radiative and radiative filaments in the northeastern part of Cygnus Loop (Vu{\v c}eti{\' c} et al.~2023). Radiative filaments are identified by their prominent [\hbox{S\,{\sc ii}}] emission, which arises from the cooled, dense post-shock gas. Vu{\v c}eti{\' c} et al.~(2023) confirmed that the radiative filaments in the northeastern rim of Cygnus Loop remnant are detected in [\hbox{S\,{\sc ii}}] observations. As they are presumed to be evolutionarily older local parts, the expectation is that the radiative filaments should have lower expansion speeds in comparison to non-radiative ones. Vu{\v c}eti{\' c} et al.~(2023) determined speeds of $39$ non-radiative and three radiative filaments. The speeds obtained for the non-radiative filaments are in the range of $220-590$\,km\,s$^{-1}$ with uncertainties of $20-50$\,km\,s$^{-1}$, mostly in agreement with previous studies. The three radiative filaments ( {group of filaments} labeled with H in Vu{\v c}eti{\' c} et al.~2023, and marked with red ellipse in Fig.~\ref{fig:1}) have significantly lower speeds of $110\rpm20$\,km\,s$^{-1}$. Given these particular velocity differences and the expected variation in particle acceleration efficiency or dominant emission mechanisms, differences in radio spectral indices between the two filament types are anticipated. High-resolution radio continuum observations can thus test whether particle acceleration (via DSA or reacceleration) operates in radiative shocks or if other processes, such as thermal bremsstrahlung, dominate. The VLA observations presented in this paper support the conclusion that synchrotron emission is not the dominant mechanism in the radiative filaments of the northeastern rim of the Cygnus Loop remnant.}

\section{VLA observations}

As the Cygnus Loop’s filaments are often only several arcseconds across, obtaining their radio spectral indices demands observations with sufficiently high resolution. Therefore, we obtained observations with VLA in both B and C configurations, at\,1 GHz  {(L-band; 1.008 -- 2.032\,GHz; total 1.024\,GHz bandwidth)} and 5 GHz  {(C-band; 4.743 -- 6.511\,GHz;  total 1.768\,GHz bandwidth)} (project 24A-190, PI D.~Urošević). The size of the field of view (FOV) at 1\,GHz was large enough to cover both radiative and non-radiative filaments with one pointing (see Fig.~\ref{fig:1}), while we observed two FOVs at 5\,GHz, each covering radiative and non-radiative filaments, as proposed by optical observations (Vu{\v{c}}eti{\'c} et al.~2023). Non-radiative filaments observed with VLA are marked with blue dashed-line ellipse ( {groups of filaments} labeled with  F and K in  Vu{\v c}eti{\' c} et al. 2023), while the three radiative filaments are marked with red dashed-line ellipse in Fig.~\ref{fig:1}  (labeled with H in  Vu{\v c}eti{\' c} et al. 2023).

We have obtained $\sim40$~hours of observations with the VLA of NRAO\footnote{The National Radio Astronomy Observatory is a facility of the National Science Foundation operated under cooperative agreement by Associated Universities, Inc.} in B and C array configurations. C array observations were done from 25$^{\rm th}$~March 2024 -- 19$^{\rm th}$~April 2024 (15 hours); { and B array observations} from 16$^{\rm th}$~August 2024 -- 3$^{\rm rd}$~September (25 hours).  {Each of these observations were split into $\sim$100\,minute  blocks, each consisting of pointings to the phase calibrator (J2052+3635), flux calibrator (3C286), and six 12 minute} exposures out of $12$ minutes on the target (filament). Total exposure time on filaments at 1\,GHz was 7.2~hours, while we had a total of 10.8~hours at 5\,GHz. 
{There are previous VLA observations which cover the same filaments observed here (Straka et al. 1986) at 17.8\,cm ($\sim$1.7\,GHz). However, due to the significantly smaller bandwidth of the older observations (37.5 MHz total) their addition is not likely to substantially improve the sensitivity of our combined 1 GHz image and so the data were therefore not incorporated.}

The data reduction was performed with the science pipeline version 2025.1.0.32~\citep{kent20}\footnote{Details of the pipeline can be found on the NRAO website: \url{https://science.nrao.edu/facilities/vla/data-processing/pipeline}} following the heuristics embedded in the pipeline.  {The measurement sets were} presented to the pipeline,   {jointly deconvolved and} self-calibrated with models derived from  {the combined dataset. Imaging was performed using the CASA TCLEAN function with Briggs weighting (robust = 0.5), multi-frequency synthesis deconvolution (convolver = mtmfs, nterms = 2), and a standard gridder with no tapering applied. The finer-than nominal angular resolution of 4.3\arcsec$\times$3.6\arcsec\ (P.A.\,=\,--84$^\circ$) for 1\,GHz and 1.6\arcsec$\times$1.4\arcsec\ (P.A.\,=\,--69$^\circ$) for 5\,GHz results from the use of robust = 0.5 weighting, which upweights the longer baselines relative to naturally weighted data. This process resulted in final 1\,GHz and 5\,GHz high-resolution images (see Fig.~\ref{fig:3}).} 

The local noise level was measured in background regions, and is around 6\, {$\mu$Jy beam$^{-1}$} for the 1\,GHz image and 2\, {$\mu$Jy\,beam$^{-1}$} for the 5\,GHz image.  {The average RMS in the residual images is reported from the pipeline as 5.6\,$\mu$Jy\,beam$^{-1}$ for 1\,GHz and 1.5\,$\mu$Jy\,beam$^{-1}$ for 5\,GHz. These values closely match the measured background noise values, indicating that the imaging is well-converged with no significant undeconvolved flux remaining.}

\section{Results and Discussion}

\begin{figure*}
  \centering
\includegraphics[width=1.5\columnwidth]{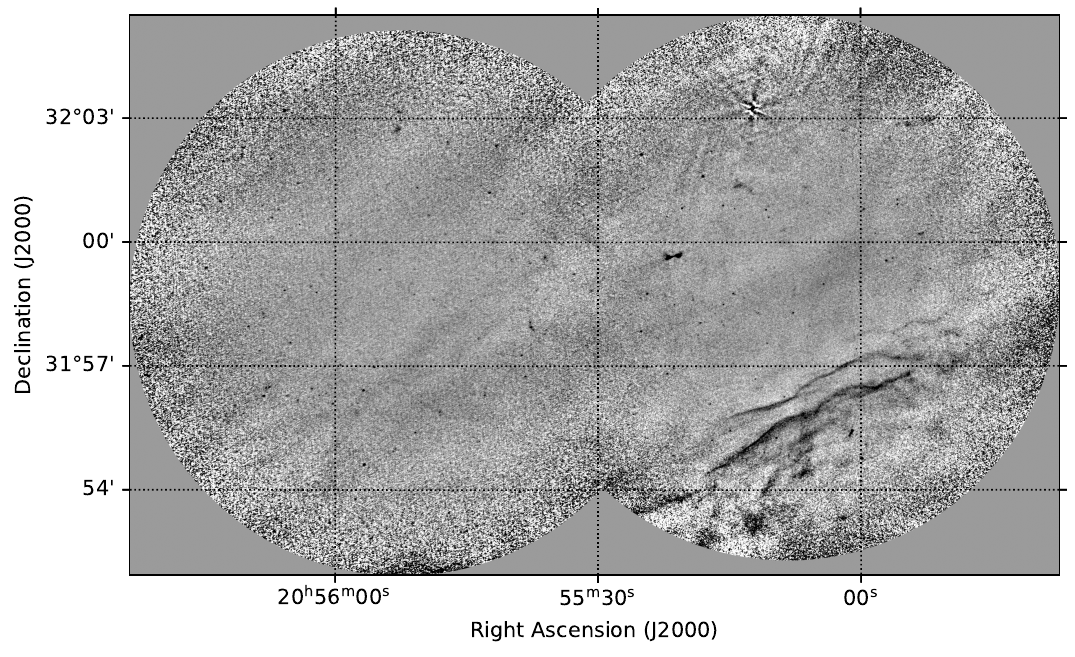}
\caption{VLA   {combined  B- and C-array image of the northeastern part of Cygnus Loop } at 5 GHz. Radiative filaments H are visible in lower right. Non-radiative filaments F and K, visible in optical emission lines (Fig.~\ref{fig:1}), are not detected in radio.}
\label{fig:3}
 \end{figure*}

In Fig.~\ref{fig:3} we present our image obtained at 5\,GHz.  {From visual inspection, it can be seen that non-radiative filaments F and K are not detected, while the radiative ones are clearly visible. The same conclusion stands also for VLA observations at 1\,GHz. We also checked the lower resolution image -- C-array at 1~GHz, for possible detection of non-radiative filaments -- but without success.} 

\begin{figure*}
  \centering
\includegraphics[width=1.7\columnwidth]{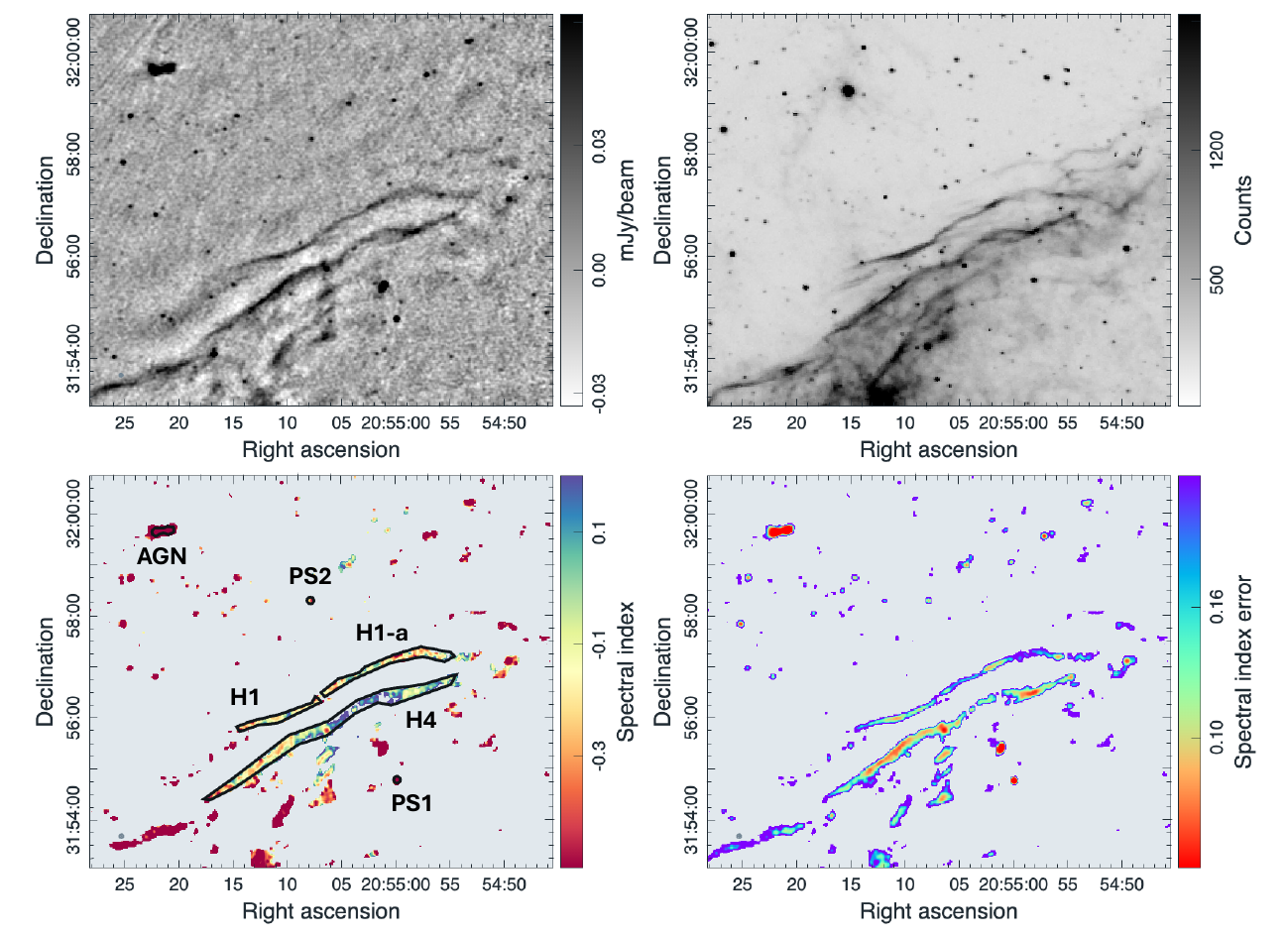}
\caption{ {1\,GHz VLA radio-continuum (top left), optical H$\alpha$ image (top right), spectral index map generated from VLA 1 and 5\,GHz observations (bottom left), and associated spectral index error map (bottom right) zoomed in on filaments of interest. Radiative filaments are divided in H1, H1-a, and H4 regions. AGN and two point sources (PS1 and PS2) are also marked. The spectral index image (bottom left) is scaled between --0.5 and 0.2 to show the filament variation. Therefore the lower end of the color bar is becoming saturated at the low end for steep spectrum sources.}}
\label{fig:6}
 \end{figure*}

 {As already noted - we use nomenclature for filaments introduced  in Vu{\v{c}}eti{\'c} et al.~(2023). In this work, localized group of filaments which was visible in ionized sulfur lines, with measured shock velocities of $\simeq 110$\,km\,s$^{-1}$, and hence taken as in radiative phase, were labeled with H1, H2 and H3. The method for measuring shock speeds in Vu{\v{c}}eti{\'c} et al.~(2023) was suitable only for nearly linear segments of filaments. Therefore, this region has much more filaments, or segments of filaments which are visible in radio, as well as in optical emission. This is why we introduced additional labeling for filaments in this region: H1-a, which continues on H1, and H4 (see bottom left panel of Fig.~\ref{fig:6}). Emission visible at the bottom left part of 1~GHz VLA image (see Fig.~\ref{fig:6},  {top} left panel), which slightly extends on the H4 filament, and shows steeper spectral index ($\alpha<-0.3$) is not labeled as it is not taken into consideration in the discussion. This segment is on the edge of the VLA field, which strongly effects spectral index measurement, and hence it is unreliable. In addition to this, region near the southern edge of 1~GHz image, is rather complex, not only filament-like, as can bee seen from optical radiation (Fig.~\ref{fig:6}, top right panel), so it could be that the foreground or background emission is contaminating filament radiation and affecting spectral index estimate.} 

{The first impression to take into account when trying to explain why we failed to detect the F and K non-radiative filaments is that they are thinner in their spatial extensions in H$\alpha$ compared to radiative filaments H (Fig.~\ref{fig:1}), and of lower optical brightness. To roughly estimate the $3\sigma$ upper limits on the integrated flux densities, we note again that the local rms noise level $\sigma_{\rm rms}$ (the continuum noise per beam), measured in source-free background regions, is $\sim5.6$\,$\mu$Jy\,beam$^{-1}$ at 1\,GHz and $\sim1.5$\,$\mu$Jy\,beam$^{-1}$ at 5\,GHz. Beam area for Gaussian synthesized beam is $17.54\ \!\mathrm{arcsec}^{2}$ and $2.54\ \!\mathrm{arcsec}^{2}$, at 1\,GHz and 5\,GHz, respectively. The angular extent of the brightest and largest non-radiative filament F, as seen in H$\alpha$ is around $\sim5-15\arcsec$ wide and $\sim140\arcsec$ long (Vu{\v c}eti{\' c} et al.~2023). Let so the filament area be of $700-2100\ \!\mathrm{arcsec}^{2}$. The number of synthesized beams covered by the filament, i.e.~the filament area divided by the effective beam area is $N_{\mathrm{beams}}$. Non-radiative filaments cover an area corresponding to roughly $N_{\mathrm{beams}}=40-276$ beams at 1\,GHz and $N_{\rm beams}=120-827$ beams at 5\,GHz, depending on the exact width and length of each filament. The estimated $3\sigma$ upper limits on the integrated flux density were therefore calculated in the standard way for extended sources as $3\sigma_{\rm rms}\times\sqrt{N_{\rm beams}}$ (Sansom et al.~2019) and amount to $\lesssim106-279$\,$\mu$Jy at 1\,GHz and $\lesssim49-129$\,$\mu$Jy at 5\,GHz, as the exact value depends on the precise area of each filament. These upper limits are lower than the measured flux densities of the radiative filaments H1, H1-a and H4 ($\sim0.9-5.6$\,mJy at 1\,GHz; see Table~\ref{tab:placeholder}). However, the low radio brightness is expected theoretically for non-radiative shocks expanding into lower-density ISM ($\sim0.2-1.5$\,cm$^{-3}$) with higher shock velocities ($\sim220-590$\,km\,s$^{-1}$), in contrast to the higher post-shock densities ($\sim6$\,cm$^{-3}$) of the radiative filaments (Raymond et al.~2020b; Tutone et al.~2021). The brightness of H$\alpha$ emission depends on the square of post-shock number density. On the other hand, the synchrotron emissivity depends only linearly on post-shock density (Bell~1978b). The non-radiative filaments, detected in H$\alpha$, emit radio synchrotron radiation \citep{uraron19}. Furthermore, the radio emission from non-radiative filaments will drop down under the sensitivity limit (in B and C configurations of VLA at 1\,GHz, and especially at 5\,GHz), much faster than H$\alpha$ emission. This is consistent with previous modeling of the Cygnus Loop (Raymond et al.~2020a; Tutone et al.~2021), which predicts a significantly lower synchrotron surface brightness in the non-radiative regions due to their lower post-shock densities.}

The spectral index map  was generated by convolving the 1 and 5\,GHz  {VLA} images to a common beam size of 5\arcsec\ and then calculating the spectral index as the linear fit of two points in log-log space using the \textsc{miriad} \textsc{maths} function (see Fig.~\ref{fig:6},  {bottom left}). We also generated a spectral index error map with the same resolution following standard error propagation.  {This error map was used to estimate the average uncertainties reported in Table~\ref{tab:placeholder}.} We marked three regions around the filaments (labeled  {H1, H1-a, and H4}), two background point sources (PS1 and PS2) and one background active galactic nucleus (AGN) (see Fig.~\ref{fig:6},  {bottom left} and Table~\ref{tab:placeholder}). The spectral index uncertainties reported in Table~\ref{tab:placeholder} are the average values from the spectral index map using the same region. For the filament regions, we also report the standard deviation of the values from the spectral index map (shown as the bracketed values in Table~\ref{tab:placeholder}, column 4), which effectively represent the variation in the spectral index value over the filament. This is a representation of how accurate the reported average value is for the entire filament, as the filaments are sufficiently large that there is spectral variation present. This background AGN is the Parkes-MIT-NRAO (PMN) source NVSS J205521+315942 (Condon et al.~1998), but it was not reported as an AGN. This was likely due to the lower resolution of PMN, and in our VLA image, it appears as a clear AGN double-lobed structure. To validate the spectral index map, we measure the flux densities of these regions individually,  {assuming a 10\% uncertainty following} the same procedure as outlined in Filipovi{\' c} et al.~(2022) and Filipovi{\' c} et al.~(2025).  {This 10\% uncertainty is a conservative estimate, which includes the nominal 3--5\% flux calibration accuracy of VLA, as well as additional contributions from residual calibration and imaging uncertainties.} We use flux measurements of these sources to calculate the spectral index independently, to compare them with the average values from the spectral index map (see Table~\ref{tab:placeholder}). We find that the manually calculated spectral indices match closely with the values obtained from the map for all points except for one of the point sources (PS2). This discrepancy is likely due to the faintness of this point source and so does not hinder the accuracy of the filament measurements, which are at least an order of magnitude brighter. The manual measurements for the filaments are consistent with those from the map, providing confidence in our measurements. The PS1, PS2, and AGN spectral indices are typical for their source type as background point sources and AGN (Filipovi{\' c} \& Tothill 2021), providing further justification.
\begin{table*}[htbp]
  \center
\begin{threeparttable}
  
      \caption{Spectral indices  {and flux density measurements} of selected regions} 
    \begin{tabular}{c|c|c|c|c}
        Region & S$_{1\,\text{GHz}}\pm\Delta S_{1\,\text{GHz}}$ & S$_{5\,\text{GHz}}\pm\Delta S_{5\,\text{GHz}}$ & $\alpha\pm\Delta\alpha$ & $\alpha$ \\
         & (mJy) & (mJy) &  (Map) & (Flux Densities)\\ \hline
         {H1-a} & 1.4$\pm$ {0.1} & 1.1$\pm$0.1 & --0.19$\pm$0.19 (0.26) & --0.16\\
         {H1} & 0.9$\pm$0.1 & 0.7$\pm$0.1 & --0.17$\pm$0.17 (0.24) & --0.16\\
         {H4} & 5.6$\pm$0.6 & 5.5$\pm$ {0.6} & --0.03$\pm$0.14 (0.19) & --0.02\\
        PS1 & 0.6$\pm$0.1 & 0.26$\pm$0.03 & --0.54$\pm$0.04 & --0.56\\
        PS2 & 0.06$\pm$0.01 & 0.05$\pm$0.01 & --0.39$\pm$0.16 & --0.11\\
        AGN & 5.9$\pm$0.6 & 1.7$\pm$0.2 & --0.79$\pm$0.02 & --0.78\\
    \end{tabular}
\begin{tablenotes}
\item  {Note: The flux density measurements assume an overall error of 10\%.} The spectral index is calculated separately from the spectral index map (column 4) and also from the flux density measurements (column 5) for validation. The spectral index uncertainties given in column 4 are taken as the average values from the spectral index error map, and the bracketed values show the standard deviation of the values in each region, reflecting the variation in the filament values.
\end{tablenotes}
   
    \label{tab:placeholder}
\end{threeparttable}
\end{table*}

The spectral index map between 1 and 5\,GHz, presented in Fig.~\ref{fig:6}  {tbottom left}, shows  {slight} differences between northern ( {H1 and H1-a} parts) and southern ( {H4} part) radiative filaments. The southern filament shows, on average, a higher  spectral index value than the northern filament. The values for these filaments are around $\alpha=0$ and $\alpha=-0.2$, respectively.  {Although the intervals of spectral index for northern and southern filament, given the large uncertainty, overlap, we interpret possible reasons for the differences.} The variation in spectral index values between these two filaments  {could be} caused by differences in the densities of the medium in both filaments. If the density of the medium is higher, the optical thickness of the medium will be higher, and it converges to the higher spectral indices of thermal bremsstrahlung emission (in the frequency region where the thermal bremsstrahlung spectra takes a break, for which the optical depth $\tau$ takes values around 1). In the limit of the totally optically thick medium, for which the optical depth $\tau\gg1$, the radio spectral index tends to a black-body radiation value of $\alpha=2$.  {The values for the spectral indices of our two filaments (southern and northern) approximately correspond to a totally thin medium where $\tau\ll1$ and so $\alpha=-0.1$.} At the end, the southern filament  {could be} at a slightly higher average medium density than the northern one, and therefore  {would have} a hig
her spectral index of thermal bremsstrahlung emission.  {Looking at the spectral index of the H1 and H4 filaments in the lower resolution data of \cite{green90} (their region C),} we see that they obtained  a steeper spectral index of $-0.35$. That might indicate a mixture of thermal and non-thermal emission,  {in this part of Cygnus Loop}. If so, it is plausible that a bright optical filaments would be dominated by thermal emission, while the non-thermal emission could be more diffuse, and therefore make a larger fractional contribution to the low-resolution observation. It is also possible that the bright optical filaments, being sheets of emitting gas seen edge-on \citep{hester87} could be optically thick, while the optically fainter, but more extended, regions surrounding the filaments, would be optically thin.

\cite{raymond20a, 2020ApJ...903....2R} and \cite{2021A&A...656A.139T} showed that the van der Laan mechanism or cosmic-ray reacceleration could match the radio surface brightness and the $\gamma$-ray emission from the radiative shock regions of the Cygnus Loop. The detection of $\gamma$-rays from the radiative shock regions of the Cygnus Loop (Tutone et al.~2021) indicates that energetic protons are interacting with the dense post-shock gas. That means that particle acceleration by the van der Laan mechanism or reacceleration does really occur, even if thermal bremsstrahlung dominates in the radio. We have shown that much of the radio continuum emission from these regions is thermal bremsstrahlung rather than synchrotron emission. The conclusions of those studies regarding the $\gamma$-ray luminosity are unchanged, but the synchrotron emission that they aimed to explain is brighter than is actually observed. Given the uncertainties in the filling factor and the pre-shock cosmic-ray electron population, and assuming that the magnetic field strength and compression ratio derived for a small area can be applied to much larger regions, their conclusion that DSA acceleration is not needed to explain the radio synchrotron emission still holds. However, our work highlights uncertainties in comparing predicted and observed non-thermal radio fluxes, so a stronger conclusion that DSA is negligible would not be warranted.

 {While the measured radio spectral indices for the radiative filaments for the northern parts H1 and H1-a, and for the southern part H4, are significantly flatter than the canonical synchrotron value of $\alpha\approx-0.5$ predicted by test-particle DSA for strong shocks, these values are fully consistent with thermal bremsstrahlung emission in the optically thin limit. The uncertainties on the spectral indices derived from the map and the standard deviation of the values within each filament region (see Table~\ref{tab:placeholder}) mean that the data cannot completely rule out a small synchrotron contribution. However, the central values and the non-detection of the non-radiative filaments~F and K strongly favor thermal bremsstrahlung as the dominant mechanism in the dense, radiative post-shock gas.

 {Weiler et al.~(1986, 1989, 2009) showed that radio emission from supernovae and young remnants frequently exhibits spectral flattening or curvature due to free-free absorption, environmental effects, and time-dependent optical depth changes. These effects are also relevant for evolutionary older SNRs expanding in dense and highly heterogeneous environments. Similarly, Dubner \& Giacani~(2015) emphasiyed the difficulties in interpreting radio spectra of evolved SNRs, where mixed thermal and non-thermal emission, incomplete frequency sampling, and limited spatial resolution often lead to apparent deviations from pure synchrotron behavior. In the specific case of the radiative shocks in the Cygnus Loop, the high post-shock densities naturally enhance the thermal bremsstrahlung contribution, making it the dominant radio continuum mechanism despite the presence of cosmic-ray electrons.}

 {We note the results of Arias et al.~(2019) for the mixed-morphology supernova remnant VRO 42.05.01, which displays a relatively flat radio spectral index at higher radio continuum frequencies ($\alpha\approx-0.37$) together with curvature that leads to spectral steepening at low frequencies (as probed by LOFAR at 143\,MHz). They emphasized that the observed curvature in the low-frequency end of the radio spectrum occurs primarily in the brightest regions of the source, while the fainter regions present a roughly constant power-law behavior between 143\,MHz and 2695\,MHz. The authors favor an explanation in which radiative shocks possess high compression ratios ($\chi>4$), such that electrons of different energies scatter over different length scales in the post-shock region and therefore sample varying effective compression ratios. Lower-energy electrons (corresponding to lower radio frequencies) probe regions closer to the shock front with lower compression, resulting in steeper local spectral indices. While the Cygnus Loop radiative filaments also exhibit flat spectral indices associated with radiative shocks, our interpretation at 1--5\,GHz differs. The observed values are much more consistent with the optically thin thermal bremsstrahlung limit ($\alpha\approx-0.1$), and the non-detection of the non-radiative filaments F and K (which would be expected to be synchrotron-bright if, e.g., DSA, Fermi II or re-acceleration were dominant) versus the clear detection of the radiative filaments H strongly favors thermal bremsstrahlung as the dominant emission mechanism. The high-compression-ratio synchrotron scenario proposed by Arias et al.~(2019) would still predict non-thermal emission, but the brightness contrast and the measured spectral indices in our VLA data support thermal dominance at these frequencies. Low-frequency observations with greater sensitivity would be needed to test whether similar curvature effects are present in the radiative shocks of the Cygnus Loop.}

\section{Summary}

In this paper, we present new VLA observations at\,1 and 5\,GHz of the selected filaments in the  {northeastern rim of Cygnus Loop}, in order to determine the radio spectral indices of non-radiative and radiative filaments previously observed in the optical range. In the observed FOVs, there are no detections of the non-radiative filaments, while the radiative filaments are clearly visible. Contrary to expected synchrotron spectral slopes of around $-0.5$, characteristic for shell-type SNRs, the observed shallower values of radio spectral indices for the radiative filaments in Cygnus Loop indicate that the thermal bremsstrahlung is the dominant radiation mechanism for the production of their radio continuum emission.

\begin{acknowledgements}
 {We thank the reviewer for  {their} comprehensive suggestions that significantly improved this article.} DU, MA and DO are supported by the Ministry of Science, Technological Development and Innovation of the Republic of Serbia, through contract no. 451-03-33/2026-03/200104. DU and MA are also supported through the joint project of the Serbian Academy of Sciences and Arts and Bulgarian Academy of Sciences - "Detection and Kinematic Characterization of Optical Counterparts to Radio Supernova Remnants".

\end{acknowledgements}





\vfill
\eject

\end{document}